\documentclass[a4paper,11pt]{article}
\pdfoutput=1 

\usepackage{jinstpub} 
\usepackage{siunitx}
\DeclareSIUnit\px{px}
\DeclareSIUnit\electron{e}
\DeclareSIUnit\dru{dru}
\graphicspath{{figures/}}

\title{\boldmath Dark Matter in CCDs at Modane (DAMIC-M): a silicon detector apparatus searching for low-energy physics processes}

\author[a,1]{S. J. Lee,\note{Corresponding author.}}
\author[a]{B. Kilminster,}
\author[a]{A. Macchiolo}

\affiliation[a]{Universit\"{a}t Z\"{u}rich,\\Winterthurerstrasse 190, 8057 Z\"{u}rich, Switzerland}
\emailAdd{stelee@physik.uzh.ch}

\abstract{Dark Matter In CCDs (DAMIC) is a silicon detector apparatus used primarily for searching for low-mass dark matter using the silicon bulk of Charge-Coupled Devices (CCDs) as targets. The silicon target within each CCD is \SI{675}{\micro\meter} thick and its top surface is divided into over 16 million \SI{15}{\micro\meter} $\times$ \SI{15}{\micro\meter} pixels. The DAMIC collaboration has installed a number of these CCDs at SNOLAB. As of 2019, DAMIC at SNOLAB has reached operational conditions with leakage current less than \SI{8.2e-22}{\ampere\per\centi\meter\squared} and a readout noise of \SI{1.6}{\electron}, achieved with 5 CCDs. A new DAMIC apparatus will be installed at Laboratoire Souterrain de Modane in a few years. The DAMIC at Modane (DAMIC-M) collaboration will be using an improved version of CCDs designed by Lawrence Berkeley National Laboratory with skipper amplifiers that use non-destructive readout with multiple-sampling, enabling the CCDs to achieve a readout noise of \SI{0.068}{\electron}. The low readout noise, in conjunction with low leakage current of these skipper CCDs, will allow DAMIC-M to observe physics processes with collisions energies as low as \SI{1}{\electronvolt}. The DAMIC-M experiment will consist of an array of 50 large-area skipper CCDs with more than 36 million pixels in each CCD. The following proceeding will introduce the DAMIC apparatus at SNOLAB and its results and as well as the capabilities and the status of the new DAMIC-M experiment.}


 \arxivnumber{2001.01209} 

\collaboration[c]{on behalf of the DAMIC and DAMIC-M collaborations}

\proceeding{15$^{\text{th}}$ Topical Seminar on Innovative Particle and Radiation Detectors (IPRD19)\\
  14-17 October 2019\\
  Siena, Italy}

\begin{document}
\maketitle
\flushbottom

\section{Introduction}
\label{sec:intro}
The DArk Matter In CCDs (DAMIC) experiment originates from Fermi National Accelerator Laboratory (Fermilab). It was originally assembled using spare Charge-Coupled Devices (CCDs) used in the Dark Energy Survey (DES) experiment\footnote{The DES experiment can be reached at \url{https://www.darkenergysurvey.org/}}. In order to use the high-resolution and low-noise scientific CCDs to search for dark matter (DM), the DAMIC collaboration has installed a shield around the devices to reduce the radioactive backgrounds. The right of \textbf{Figure \ref{fig:CCD Diagram}} illustrates a layout of a DAMIC experiment installed in the Sudbury Neutrino Observatory Laboratory (SNOLAB) \cite{Chavarria:2014ika}. The experiment was installed underground to take advantage of \SI{2}{\kilo\meter} of rock (~\SI{6}{\kilo\meter} water equivalent) to reduce the cosmogenic particles. This experiment is called DAMIC at SNOLAB (DAMIC-SNOLAB).
\subsection{DAMIC at SNOLAB}
\begin{figure}[htbp!]
\centering 
\includegraphics[width=.8\textwidth,origin=c ]{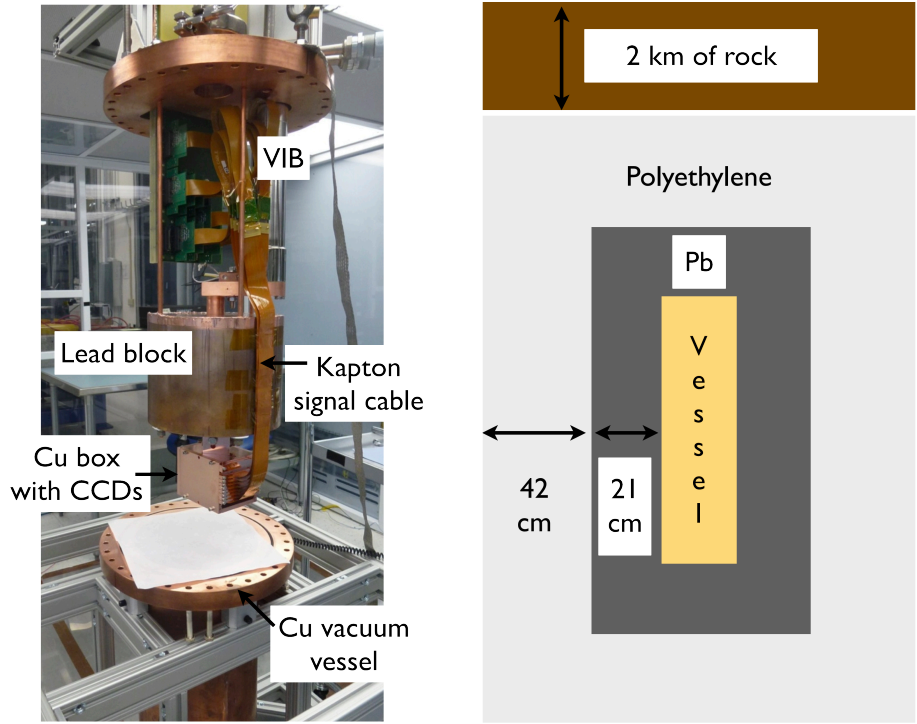}
\caption{\label{fig:CCD Diagram} Left shows the DAMIC-SNOLAB experiment without its external shields and right shows a diagram demonstrating the configuration of external shield that surrounds the experiment.}
\end{figure}

In addition to the natural shielding provided by the rocks at SNOLAB, DAMIC-SNOLAB has an additional \SI{42}{\centi\meter} of polyethylene and \SI{21}{\centi\meter} of lead surrounding the experiment to shield the CCDs from secondary particles from cosmogenic sources and surrounding radioactive background sources. Inside a copper vessel containing the CCDs, a selected number of CCDs are further surrounded by radio-pure ancient lead to provide more shielding from local radioactive sources. The left of \textbf{Figure \ref{fig:CCD Diagram}} shows DAMIC-SNOLAB without all of its shielding. Within the copper box there are currently 8 specialized CCDs produced by the Lawrence Berkeley National Laboratory (LBNL) and Teledyne DALSA\footnote{Teledyne Dalsa can be reached at \url{https://www.teledynedalsa.com/en/home/}}.

\subsection{CCDs for DAMIC}
Since the original DAMIC experiment at Fermilab, the DAMIC-SNOLAB collaboration has modified several components of the LBNL CCDs. Some of the modifications include increasing the thickness of the CCD, removal of reflection layer in the backside of the CCD (which was a requirement to increase sensitivity to the infrared spectrum for the DES experiment) and radio-purity and resistance of the bulk. A proper classification of the CCDs used by the DAMIC collaboration is large area, thick, 3-stage high voltage compatible, p channel in n bulk, fully-depleted back-illuminated scientific grade CCDs. Each of the CCDs installed in SNOLAB is \SI{675}{\micro\meter} thick, weighs \SI{6.0}{\gram} and is comprised of over 16 million pixels and with each pixel covering an area of \SI{15}{\micro\meter} $\times$ \SI{15}{\micro\meter}. These CCDs are fully depleted at \SI{40}{\volt} and kept at a pressure of \SI{e-7}{\milli\bar} and a temperature of \SI{135}{\kelvin} \cite{ccdspec}.
\subsection{Scientific CCDs}
It is worth noting that these CCDs are scientific-grade CCDs and therefore have an advantage that the operation and configuration of each exposure or image taken can be controlled. This includes power variables such as bias voltages, drain voltage, and reset voltage, as well as the clock voltages used for holding charges within each pixel during the exposure, and the clock sequences for reading out the charges in each pixel.

\begin{figure}[htbp]
\centering 
\includegraphics[width=1\textwidth,origin=c ]{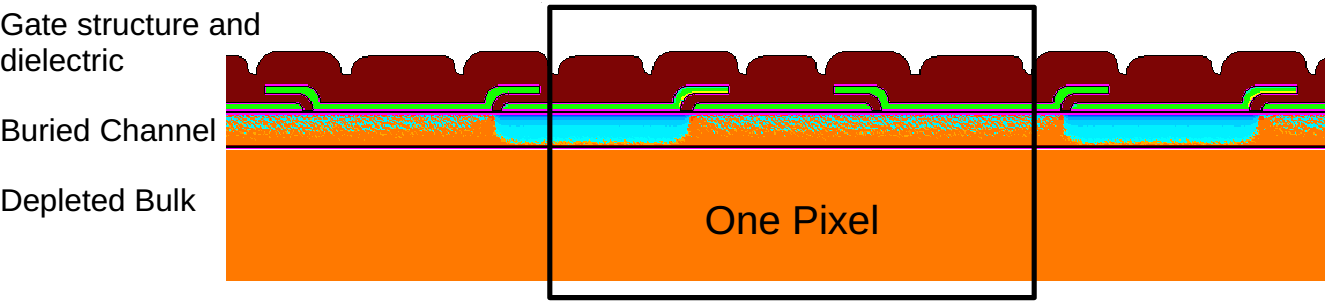}
\caption{\label{fig:CCD_TCAD} A TCAD-generated illustration of top structure of CCDs used by DAMIC during charge transfer along its buried p channel.}
\end{figure}

\textbf{Figure \ref{fig:CCD_TCAD}} exemplifies the general top surface structure of an individual pixel in these CCDs generated using Synopsys Sentarus TCAD\footnote{More information on TCAD tools can be found at \url{https://www.synopsys.com/silicon/tcad.html}}. The top brown layer is an oxide passivation layer. Some of the brown oxide layer is thermally grown in between green polysilicon gate structures. The n doped polysilicon gate structures are in electrical contact with a buried p channel through a dielectric junction made out of a combination of SiO$_2$ and Si$_3$N$_4$. Under the buried p channel is the bulk of the CCD. Approaching the backside of the CCD, the bulk is negatively doped until reaching in-situ doped polysilicon (ISDP) deposition layer, and passivated with alternating layers of SiO$_2$ and polysilicon.

\textbf{Figure \ref{fig:CCD_TCAD}} also demonstrates the charge flow and collection in the top structure of the CCDs. There are 3 different geometries of polysilicon gate structure. Starting at the leftmost component of the figure are gate structures 1, 2, 3 and repeating. These are the 3-stage gate structures. Gate structure 1 is defined as the center of one pixel and each pixel shares the borders where gate structures 2, and 3 overlap as seen in the black rectangle. In addition, the figure illustrates in teal a large concentration of charge carriers (holes) being localized to under a single gate structure (gate 3). This TCAD simulation demonstrates the charge transfer between each gate structure. Similar simulations can be used to optimize the charge collection and transfer efficiency.

\section{Results from DAMIC at SNOLAB}\label{sec:damicsnolab}
DAMIC-SNOLAB has been operational for over 6 years searching for DM. During the search, the collaboration has seen some standard model particle tracks, and radio-impurities from within the shielding, cables and CCDs themselves. The collaboration has been able to model the depth of the standard model (SM) particle tracks, set constraints on some DM candidates, model the radio-impurity of detector components and, in addition have been able to optimize the operating parameters of the CCDs. DAMIC-SNOLAB CCDs have been operating with a leakage current \SI{2e-22}{\ampere\per\centi\meter\squared} with a readout noise equivalent to \SI{1.6}{\electron}\footnote{\SI{}{ \electron} denotes electron charges}.
\subsection{Particle tracks in CCDs}
\begin{figure}[htbp]
\centering 
\includegraphics[width=1\textwidth,origin=c ]{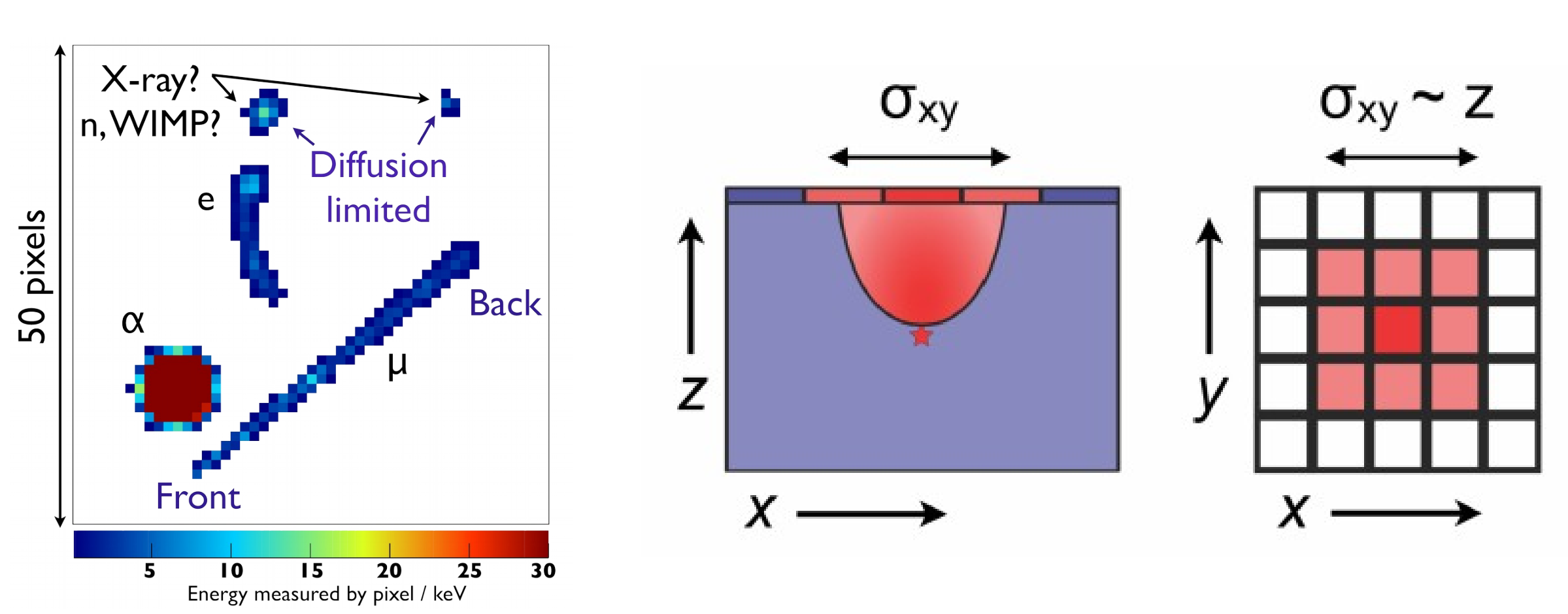}
\caption{\label{fig:Particle_Tracks} The left shows particle tracks from alpha, muon, electron and diffusion-limited x-ray tracks with relative scale. The center and right show a diagram illustrating the diffusion of charge carriers as a function of depth resulting in diffused particle tracks \cite{Chavarria:2014ika}.}
\end{figure}

While CCDs are 2 dimensional imaging devices, each CCD can be used to reconstruct 3 dimensional particle tracks. As incident particles ionize and produce excess charge carriers while the device is fully depleted, the charge carriers must travel up to the top polysilicon gate structure in order to be collected. As previously stated the CCDs installed at SNOLAB are \SI{675}{\micro\meter} thick and fully depleted at \SI{40}{\volt} and are held at \SI{135}{\kelvin}. This leaves some distance and time before charge carriers reach the gate structure, allowing some of the charge carriers to diffuse into adjacent pixel structures. \textbf{Figure \ref{fig:Particle_Tracks}} illustrates a simplified diagram of charge diffusion to the right. The left of \textbf{Figure \ref{fig:Particle_Tracks}} shows the actual particle tracks left by some SM particles. As long as the CCD is not overdepleted, the diffusion of charge carriers can be modelled using the Point Spread Function (PSF) and can be used to determine the depth of particle tracks within the bulk \cite{janesick,PSF,hollandFD,hollandHV}.

\subsection{Search for dark matter candidates}
\begin{figure}[htbp]
\centering 
\includegraphics[width=1\textwidth,origin=c ]{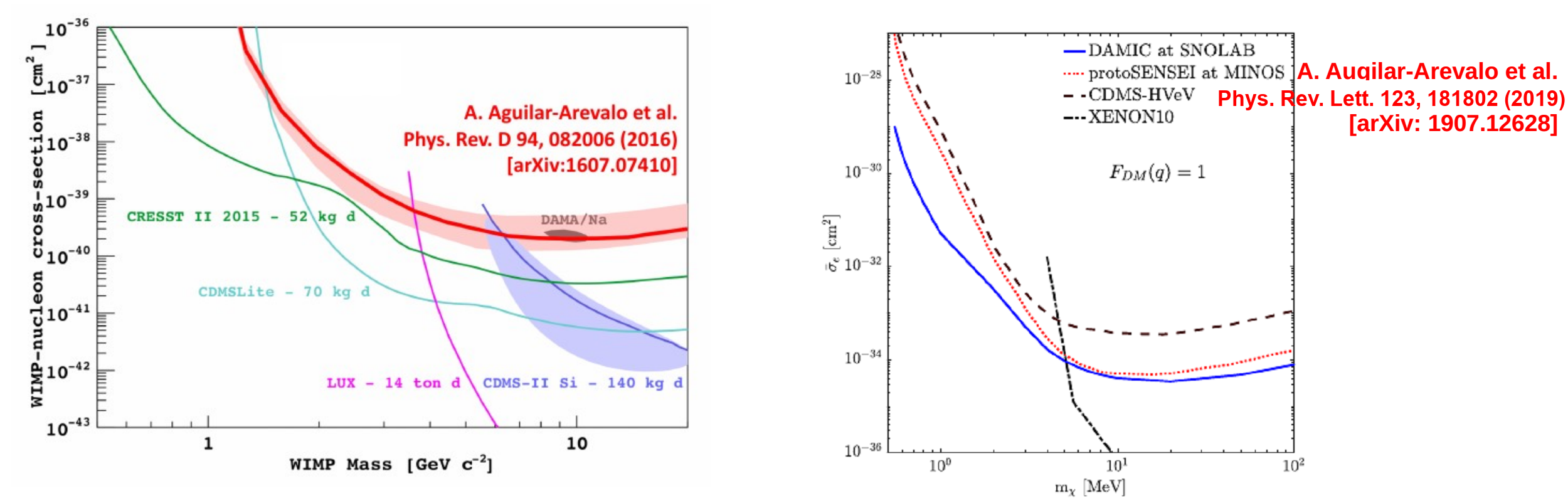}
\caption{\label{fig:DM_Results} Left shows upper limit on the cross section and mass of  Weakly Interacting Massive Particles (WIMP) bold red line demonstrates the constraints set by the DAMIC-SNOLAB experiment \cite{lowmasswimppaper}. Right shows constraints of the cross section and mass of hidden photon interacting with standard model electron set by the DAMIC-SNOLAB experiment \cite{hiddenphotonpaper}. }
\end{figure}

By modelling PSF and optimizing the operating parameters of the CCDs using SM particle interactions, DAMIC-SNOLAB has been able to search for some of the DM candidates. Two primary DM candidates searched for at DAMIC-SNOLAB are low mass, Weakly Interacting Massive Particles (WIMPs) and hidden photons (often denoted as $\gamma_{\chi}$). Some of the latest results can be found in \textbf{Figure \ref{fig:DM_Results}} and in \cite{lowmasswimppaper,hiddenphotonpaper}.

\subsection{Background model}\label{sec:background}
One of the key challenges of searches for dark matter is to eliminate the flux of SM particles. The SM particles can be introduced into a DM experiment by minuscule amounts of radioactive contamination. The unit of background in DM experiments is typically the \SI{}{\dru} (\SI{1}{Event\per\kilo\electronvolt\per\kilo\gram\per\day}\footnote{\SI{}{\kilo\gram} here should not be confused with \SI{}{\kilo\gray} and \SI{}{\day} denotes day }). Obvious large contributors to these radioactive sources are naturally occuring unstable radon, and radioactive isotopes in lead such as $^{210}$Pb. Radon can be reduced through a radon-free, overpressure environment, typically using nitrogen gas, and by selecting ancient lead, with reduced isotope levels. However there are other smaller background contributors that together make a large impact to a DM experiment. Almost all of the components that make up the DAMIC-SNOLAB experiment have been tested for their radioactivity in destructive tests such as Inductively Coupled Plasma-Mass-Spectrometry (ICP-MS). Using the results of the ICP-MS of all of the materials used to construct the detector, the radioactivity at the position of each sensor can be simulated using GEANT 4. Combining the lab tests results and simulations, DAMIC-SNOLAB currently has achieved a background of \SI{11.8}{\dru}. This can further be improved by increasing or decreasing the detector material, depending on the radio-purity of the materials used.
\section{DAMIC at Modane}
As stated in \textbf{Section \ref{sec:damicsnolab}}, DAMIC-SNOLAB has been operating for over 6 years. To achieve better sensitivity to dark matter, the experiment would need to achieve lower background rates and increase the active detector mass. As of 2018, a new collaboration called DAMIC at Modane (DAMIC-M) has been formed to install a new DAMIC experiment at Laboratoire Souterrain de Modane (LSM). At LSM, The DAMIC-M collaboration plans to install new larger CCDs with lower background rates and improved electronics, along with improvements to the design of the experiment\footnote{The DAMIC-M collaboration can be reached at \url{https://damic.uchicago.edu/}}.
\subsection{Preliminary design of DAMIC-M}
One of the limiting factors for the DAMIC-SNOLAB was the active cooling and heating elements and the  vacuum pump, which generated electrical noise. As the CCDs are operated in vacuum at a pressure of \SI{e-7}{\milli\bar}, the cooling element must be in thermal and electrical contact with each device, introducing a challenge for providing a common device and earth ground. Furthermore, since the experiment is underground, the definition of common earth ground is not well defined. Without a proper earth ground, the motors used in vacuum pump and cryocooler generate minuscule amount of excess charge that produces multiple artifacts in CCDs. DAMIC-M will be designed to hold vacuum passively using charcoal cryo-pump and will be using liquid nitrogen to cool its sensors. This will reduce the possibility of introducing electrical noise into the sensors.
\subsection{Reducing the background}
As mentioned in \textbf{Section \ref{sec:background}}, DAMIC-SNOLAB had many well-studied small background sources. Some of these can be reduced by simply reducing the amount material used and also by a careful production of the detector material. First, all of the components of DAMIC-M will be tested in laboratory for their radioimpurity. Second, all of these components will be transported and produced under some form of controlled shielding from cosmogenic particles.

The most crucial detector elements are obviously the CCDs. The silicon ingot used for producing the CCDs is purchased from Topsil\footnote{Topsil can be reached at \url{http://www.topsil.com/}} located in Scandinavia. This ingot will be processed into wafers and transported first to the east coast and then the west coast of North America. Once unshielded, the silicon ingot is expected to be exposed to cosmogenic particles at an unacceptable level during transportation. Therefore, a specially designed sea container providing shielding from cosmogenic particles will be used to transport the silicon components that will eventually become CCDs. Furthermore, the CCDs will be produced under shielding further reducing the exposure to cosmogenic particles. In addition, the copper components of the detector will be made using electroforming techniques in an underground laboratory and similar care will be taken during transportation. With such care in production and transportation of the detector components, DAMIC-M aims to have background levels in the CCDs of \SI{0.1}{\dru}.
\subsection{The skipper CCDs}
Since the installation of DAMIC-SNOLAB, LBNL has been able to produce a skipper amplifier for their CCDs \cite{skipperamp}. The CCDs equipped with these amplifiers are called skipper CCDs. The special feature of the skipper amplifier is that the output node of the readout structure is floating rather than being in constant contact with the rest of the readout structure. Once the charge reaches the output node, it is either collected into the readout electronics, or discarded through the drain. By being able to break the contact to the output node, the charge can skip the output node and be sampled multiple times without being destroyed before being readout \cite{skipperamp,janesick}. 
\begin{figure}[htbp]
\centering 
\includegraphics[width=1\textwidth,origin=c ]{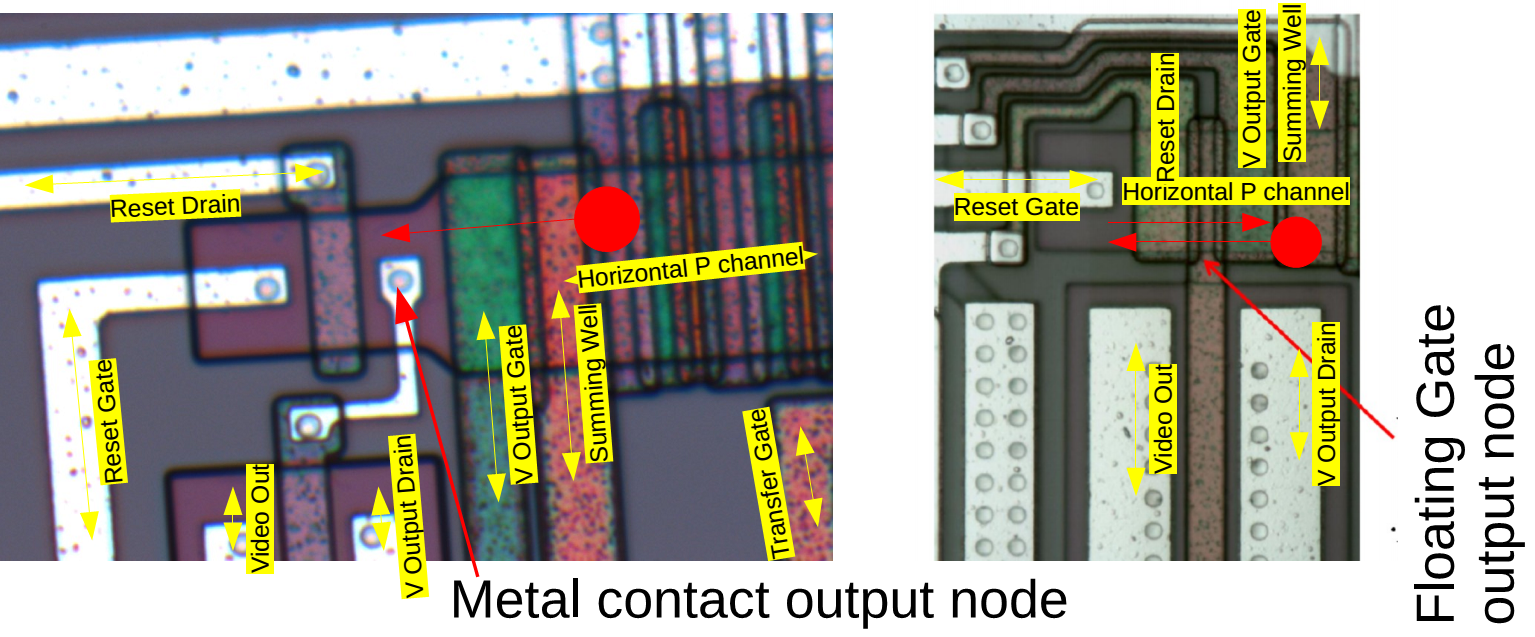}
\caption{\label{fig:Amplifiers} Left shows a conventional amplifier structure of a 2004 LBNL CCD seen from the top surface. Right shows an image of skipper amplifier structure seen from top surface from \cite{skipperamp} with permission.}
\end{figure}
\textbf{Figure \ref{fig:Amplifiers}} shows the surface structure of both conventional and skipper amplifiers. Most of the CCD top surface structures have been labelled. The red circle illustrates the path of the charge. In a conventional amplifier, the charge can only be sent in one direction, towards the output node. For a CCD with a skipper amplifier, a single packet of charge (charge collected from a single pixel) can enter the video output gate, while the next packet is held behind the summing well. From this point on, a packet of charge can be sampled multiple times skipping the output node. Once the packet enters the output node, the charges are finally read or destroyed. By being able to perform non-destructive sampling of a single charge packet multiple times, readout noise can be reduced.

\begin{figure}[htbp]
\centering 
\includegraphics[width=0.5\textwidth,origin=c ]{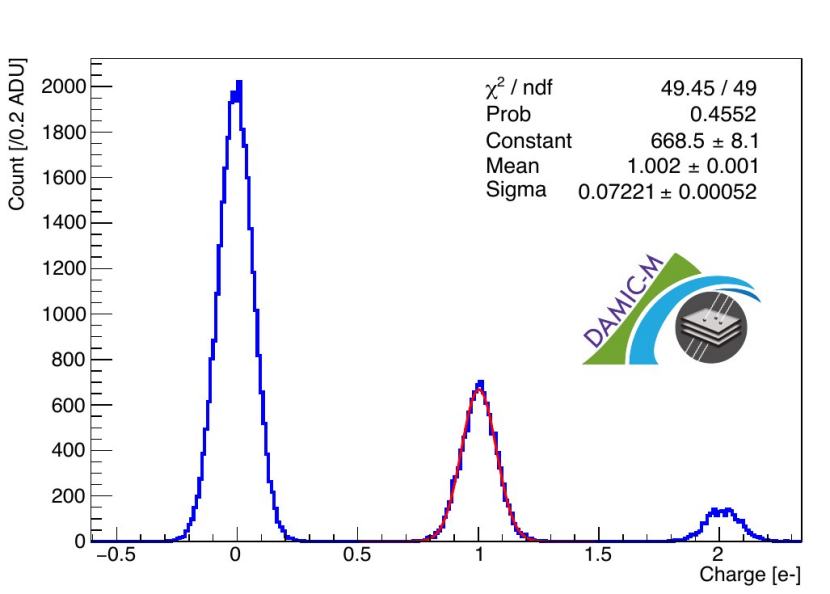}
\caption{\label{fig:single_electron_resolution} Readout performed by DAMIC-M collaboration using a 1k x 6k prototype skipper CCD to demonstrate sub-electron resolution readout noise. Here, the readout noise is equivalent to (\SI{7.22}{}$\pm$\SI{0.05}{})\SI{e-2}{\electron}.}
\end{figure}

The DAMIC-M collaboration has been operating the prototype skipper CCDs on test stands at above ground laboratories, and has been able to read out charges from prototype skipper CCDs with readout noise less than 1 electron as demonstrated in \textbf{Figure \ref{fig:single_electron_resolution}}. DAMIC-M intends to install 50 skipper CCDs, with each CCD having over 36 million \SI{15}{\micro\meter} $\times$ \SI{15}{\micro\meter} pixels.

\subsection{Simulations}
Since all of the components of the DAMIC-M detector will be assayed for their radioactivity, the radioactivity of the detector can be simulated using GEANT4. As a result, the amount of radiation reaching the CCDs will be simulated. These types of simulations will be used to determine the optimal sizes of shielding to maximize the shielding from cosmogenic and other local radioactive sources but with minimal radioactivity within the shielding materials.

DAMIC-M will also be using Synopsys Sentarus TCAD to optimize the design of the new skipper CCDs and the operating parameters. TCAD will also be used to better model the PSF and improve the accuracy of reconstructed particle tracks within the CCDs.
\section{Conclusion}
DAMIC-SNOLAB has been operating up to eight \SI{16}{\mega\px} CCDs at \SI{135}{\kelvin}, in vacuum at a pressure of \SI{e-7}{\milli\bar}, fully depleted at \SI{40}{\volt} with leakage current down to \SI{2e-22}{\ampere\per\centi\meter\squared}, readout noise equivalent to \SI{1.6}{\electron} and with radioactive background of \SI{11.8}{\dru}.

Since the installation of DAMIC-SNOLAB, technology and techniques for developing and operating DM detectors have significantly improved, and a new collaboration called DAMIC-M has been formed to install a new DAMIC detector setup at Laboratoire Souterrain de Modane. DAMIC-M has since become a CERN recognized experiment and will also search for displacement damage at atomic scale and model low energy non-ionizing energy loss in collaboration with the RD50 collaboration at CERN\footnote{The RD50 collboration can be reached at \url{https://rd50.web.cern.ch/rd50/}}.

In 2020, a proof-of-concept prototype of DAMIC-M will be installed. The final detector design of DAMIC-M will have 50 skipper CCDs with readout noise less than (\SI{7.22}{}$\pm$\SI{0.05}{})\SI{e-2}{\electron} in a charcoal cryo-pump with background noise lower than \SI{0.1}{\dru}.

%
%
%
%
%
%
%


\acknowledgments
The author would like to acknowledge the Swiss National Science Foundation (SNSF) and their Funding LArge international REsearch projects (FLARE) grant for enabling this research.

The author would like to also acknowledge S. E. Holland and C. J. Bebek from LBNL for their development of their CCDs and correspondence which has been supporting research for both DAMIC-SNOLAB and DAMIC-Modane.

The author would also like to acknowledge Sub-Electron-Noise SkipperCCD Experiment Instrument (SENSEI) and Fermilab for their work together with LBNL developing sub-electron noise skipper CCDs.

The author would also like to acknowledge American Association of Variable Star Observers (AAVSO) and their CEO Dr. Stella Kafka for writing and maintaining a wonderful comprehensible reference resource for CCD photometry.

The author would also finally like to acknowledge the extended members of the DAMIC-SNOLAB and DAMIC-M collaborations.



\begin{thebibliography}{99}
\bibitem{Chavarria:2014ika} 
  A.~E.~Chavarria {\it et al.},
  Phys.\ Procedia {\bf 61}, 21 (2015)
  doi:10.1016/j.phpro.2014.12.006
  [arXiv:1407.0347 [physics.ins-det]].
  
\bibitem{ccdspec} 
C. Bebek and N. Roe, \emph{4k x 2k and 4k x 4k CCD Users Manual}
Rev. 3b March 23, 2011

\bibitem{janesick}
James R. Janesick, \emph{Scientific Charge-Coupled Devices},
ISBN:9780819436986, Volume PM83, 15 January 2001.

\bibitem{PSF}
S.E. Holland et al., \emph{Point-spread function in depleted and partially depleted CCDs},
ODT 99 LBNL-45276. 15 September 1999.

\bibitem{hollandHV}
S.E. Holland et al., \emph{High-voltage-compatible, fully depleted CCDs},
Lawrence Berkeley National Laboratory, 2006-05-15 \url{https://escholarship.org/uc/item/6ft833qz}

\bibitem{hollandFD}
  S.~E.~Holland, C.~J.~Bebek, W.~F.~Kolbe and J.~S.~Lee,
  JINST {\bf 9} (2014) C03057
  doi:10.1088/1748-0221/9/03/C03057
  [arXiv:1403.6185 [astro-ph.IM]].

\bibitem{hiddenphotonpaper}
  A.~Aguilar-Arevalo {\it et al.} [DAMIC Collaboration],
  Phys.\ Rev.\ Lett.\  {\bf 123} (2019) no.18,  181802
  doi:10.1103/PhysRevLett.123.181802
  [arXiv:1907.12628 [astro-ph.CO]].

\bibitem{lowmasswimppaper}
  A.~Aguilar-Arevalo {\it et al.} [DAMIC Collaboration],
  Phys.\ Rev.\ D {\bf 94} (2016) no.8,  082006
  doi:10.1103/PhysRevD.94.082006
  [arXiv:1607.07410 [astro-ph.CO]].

\bibitem{skipperamp}
  C.~J.~Bebek {\it et al.},
  Proc.\ SPIE Int.\ Soc.\ Opt.\ Eng.\  {\bf 8453} (2012) 845305.
  doi:10.1117/12.926606










\end{thebibliography}
\end{document}